\pdfoutput=1

\documentclass[conference]{IEEEtran}

\usepackage{cite}

\usepackage{tikz}
\usepackage{pgfplotstable}
\pgfplotsset{compat=1.14}
\usepackage{subfiles}
\usepackage{textcomp}
\usepackage{RobStd}
\usepackage{graphicx}
\usepackage[font=small]{caption}
\usepackage[font=footnotesize]{subcaption}
\ifCLASSINFOpdf
\else
\fi

\usepackage{amsmath,amssymb,amsfonts}

\usepackage{algorithmic}

\begin{document}
%

\title{A Hybrid FeMFET-CMOS Analog Synapse Circuit for Neural Network Training and Inference}

\author{Arman Kazemi$^{*}$, Ramin Rajaei$^{*}$, Kai Ni$^\dagger$, Suman Datta$^{*}$, Michael Niemier$^{*}$, X. Sharon Hu$^{*}$\\
$^{*}$University of Notre Dame, $^\dagger$Rochester Institute of Technology
\vspace{-0.2in}
}

\maketitle

\begin{abstract}
An analog synapse circuit based on ferroelectric-metal field-effect transistors is proposed, that offers 6-bit weight precision. The circuit is comprised of volatile least significant bits (LSBs) used solely during training, and non-volatile most significant bits (MSBs) used for both training and inference. The design works at a 1.8V logic-compatible voltage, provides 10\textsuperscript{10} endurance cycles, and requires only 250ps update pulses. A variant of LeNet trained with the proposed synapse achieves 98.2\% accuracy on MNIST, which is only 0.4\% lower than an ideal implementation of the same network with the same bit precision. Furthermore, the proposed synapse offers improvements of up to 26\% in area, 44.8\% in leakage power, 16.7\% in LSB update pulse duration, and two orders of magnitude in endurance cycles, when compared to state-of-the-art hybrid synaptic circuits. Our proposed synapse can be extended to an 8-bit design, enabling a VGG-like network to achieve 88.8\% accuracy on CIFAR-10 (only 0.8\% lower than an ideal implementation of the same network).
\end{abstract}

\IEEEpeerreviewmaketitle

\section{Introduction}

Given the exponential growth of data, researchers are investigating new ways to automate data analysis through the use of deep neural networks (DNNs).  
DNN accelerators that perform multiplication and addition in the analog domain, e.g., using resistive devices as synapses in crossbar arrays, are appealing and could reduce the time and energy associated with DNN training and inference~\cite{han2016eie} by orders of magnitude~\cite{gokmen2016acceleration}. 

Per~\cite{gokmen2016acceleration}, to offer the greatest application-level impact, synapses (i.e., crosspoints in crossbar arrays) should afford \textbf{(i)} update pulses with 1ns width and \textpm1V magnitude for potentiation and depression (i.e. increasing and decreasing conductance, respectively), and \textbf{(ii)} symmetric and linear weight updates where weights have 1000 unique states/offer \texttildelow10-bit precision. Emerging non-volatile resistive devices, e.g., resistive RAM~\cite{woo2016improved}, ferroelectric FETs (FeFETs)~\cite{jerry2018ferroelectric}, and phase-change memory (PCM)~\cite{ambrogio2018equivalent} are the primary candidates for crosspoint synapses within a crossbar array, due to their lower area and higher density when compared to their CMOS counterparts. However, crossbar arrays comprised of emerging devices cannot deliver comparable training/inference accuracy as software implementations of the same network due to their non-linear, asymmetric weight updates~\cite{chen2018technological}.

Alternatively, CMOS-based synapses (e.g.~\cite{kim2017analog}) offer high linearity and symmetry with rapid updates but at the expense of lower density, higher energy, and volatility. To exploit the benefits of both CMOS and emerging devices, hybrid synaptic circuits built with both CMOS and some emerging devices have been introduced (e.g.,~\cite{ambrogio2018equivalent,sun2018exploiting}). However, these designs incur high peripheral circuitry and delay overhead, require high write voltage, and/or have low endurance.

We propose a hybrid, high precision synapse circuit comprised of ferroelectric metal field-effect transistors (FeMFETs)~\cite{ni2018soc} and CMOS transistors. FeMFETs represent non-volatile most significant bits (MSBs) and are used during training and inference. CMOS devices represent volatile least significant bits (LSBs) and are only employed during training. The proposed synapse works at a logic-compatible voltage of 1.8V, requires symmetric and identical 250ps programming pulses for very fast potentiation and depression, and provides $10^{10}$ MSB endurance cycles. The synapse circuit is simulated using an experimentally calibrated FeMFET model~\cite{ni2018soc} and a 65nm CMOS PTM~\cite{cao2009predictive} model (for uniform comparisons to other approaches). When training a variant of LeNet~\cite{lecun2015lenet} on the MNIST~\cite{lecun-mnisthandwrittendigit-2010} dataset, we achieve an accuracy of 98.2\%, which is only 0.4\% lower than an ideal implementation of the same network with the same bit precision. Furthermore, the proposed synapse offers improvements of up to 26\% in area, 44.8\% in leakage power, 16.7\% in LSB update pulse duration, and two orders of magnitude in endurance cycles, when compared to state-of-the-art synaptic circuits. The synapse design is extendable to an 8-bit design by employing an extra MSB device. When training a VGG-like network with the CIFAR-10 dataset~\cite{krizhevsky2009learning} and the 8-bit extended synapse, we achieve a classification accuracy of 88.8\% (0.8\% lower than an ideal implementation of the same network). 
\begin{figure}[t]
        \centerline{\includegraphics[width=.8\columnwidth]{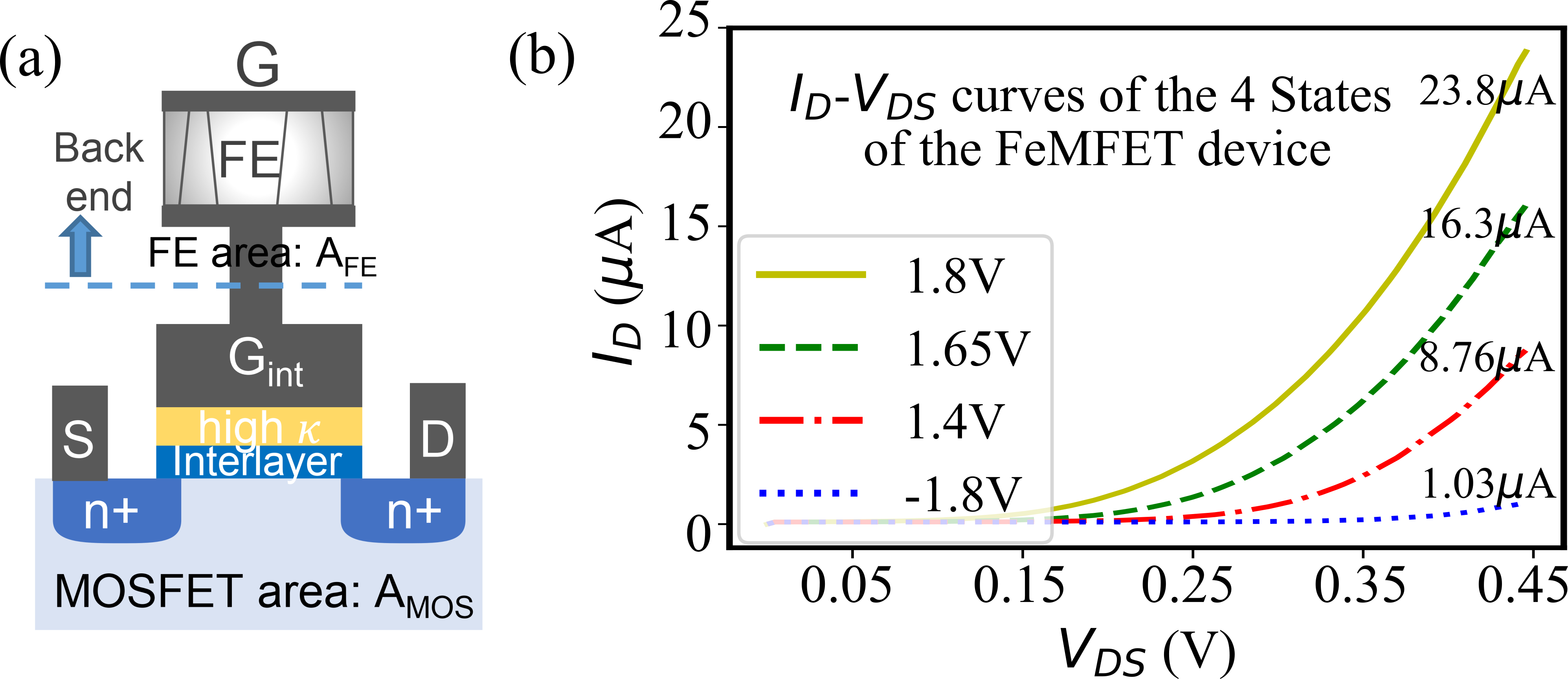}}
        \caption{{\bf (a)} FeMFET structure including a MOSFET and a back-end MFM capacitor (from~\cite{ni2018soc}); {\bf (b)} $I_D$-$V_{DS}$ curve showing four FeMFET states.}
        \label{fig:femfet}
        \vspace{-0.2in}
\end{figure}

\section{Background and Related Work}
\label{sec:bg}
\subsection{The FeMFET device}

A FeMFET incorporates a ferroelectric (FE) capacitor in the {\em back-end of line (BEOL)\/} (Fig.~\ref{fig:femfet}(a)), which reduces the maximum required programming voltage to a logic-compatible level of 1.8V (compared to 4V in FeFETs) and increases endurance to $10^{10}$ cycles. These improvements are obtained by independently optimizing the area of the FE capacitor and the MOSFET, which allows for maximum voltage drop across the FE~\cite{ni2018soc}. FeMFETs have been experimentally demonstrated~\cite{ni2018soc}. We adapt a model, calibrated by experimental data, to represent the characteristics of FeMFETs~\cite{ni2018circuit}. Fig.~\ref{fig:femfet}(b) shows four different states (i.e., on current) of a FeMFET obtainable with different programming voltages. While the FeMFET device does not in and of itself deliver 1000 states~\cite{gokmen2016acceleration}, its low write voltage and high endurance make it attractive as a synaptic device for crossbar arrays/hybrid synaptic circuits.

\begin{figure}[t]
        \centerline{\includegraphics[width=.7\columnwidth]{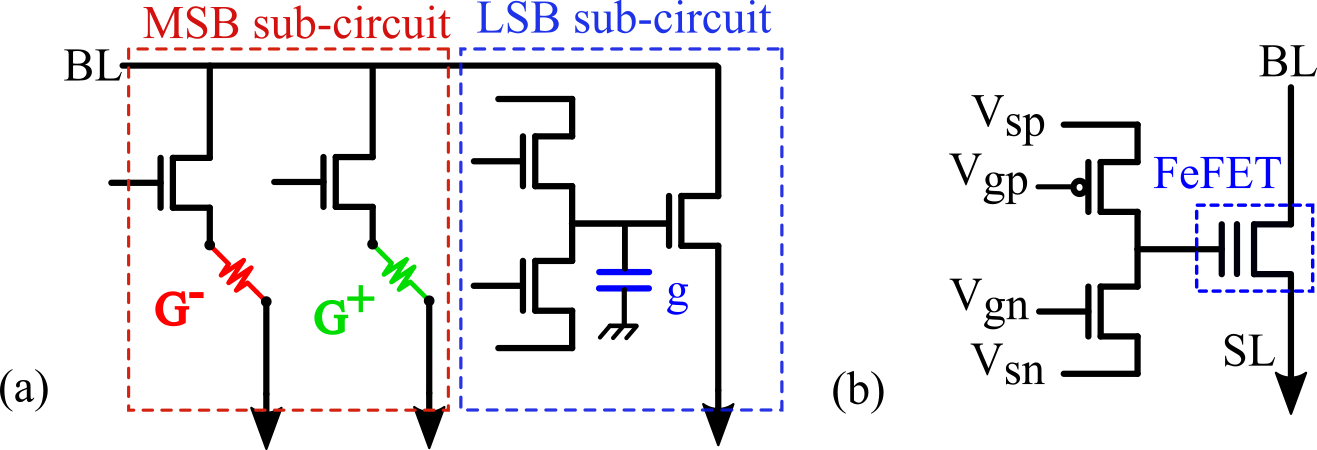}}
        \caption{State-of-the-art hybrid synapse circuits: {\bf (a)} The synapse circuit in~\cite{ambrogio2018equivalent} uses 2 PCM devices as MSBs and a CMOS sub-circuit as LSBs; {\bf (b)} The synapse circuit proposed in~\cite{sun2018exploiting} where polarization states of an FeFET represent the MSBs and the gate voltage of the FeFET represents the LSBs.}\label{fig:prior_work}
        \vspace{-0.25in}
\end{figure}

\subsection{Existing hybrid synapse circuits}
Hybrid synapse circuits built with CMOS and emerging devices can exploit the advantages of both types of devices. A hybrid synapse was first proposed in~\cite{ambrogio2018equivalent} where two PCM devices (for positive and negative values) represent the MSBs, and a three-transistor plus one capacitor (3T1C) circuit represents the LSBs (Fig.\ref{fig:prior_work}(a)) -- referred to as a 2PCM+3T1C design. The MSBs are non-volatile and used for training and inference, while the LSBs are volatile and are only used during training (i.e., as higher precision is required for training than inference~\cite{ambrogio2018equivalent,sun2018exploiting}). This design requires update pulses of 300ps width and 1V magnitude. It also requires a 3-phase read-out which induces additional delay. Furthermore, once the three values ($G^+$, $G^-$, and $g$ in Fig.~\ref{fig:prior_work}(a)) are read from the synapse and digitized, the actual contribution of the synapse to the output must be calculated as $F\times (G^+-G^-) + g$, where \textit{F} is the gain factor. This operation is typically done with multiplication in the periphery of the crossbar. Training with this synapse structure on the MNIST dataset achieved an accuracy of 97.95\% for an MLP with 784-150-125-10 neurons. Note that two PCM devices are required in this design as PCM devices do not have bi-directional symmetry in weight updates, which inversely impacts crossbar array area and energy.

In~\cite{sun2018exploiting}, a 2T-1FeFET (2T1F) synapse circuit (Fig.\ref{fig:prior_work}(b)) was proposed and can obtain 6 or 7 bits of precision via 2 non-volatile MSBs represented by the polarization states of an FeFET, and 4 or 5 volatile LSBs represented by the gate voltage of the FeFET. This design only uses 3 devices and works with a single-phase read-out scheme. The 2T1F synapse achieves a training accuracy of 97.3\% and 87\% for a variant of LeNet and a VGG-like network for the MNIST and CIFAR-10 datasets, respectively. However, this circuit uses I/O transistors with 3.3V supplies and an FeFET with programming voltages in the 2V-4V range. Collectively, these requirements increase power consumption and complicate the logic compatible implementation of the 2T1F design. Furthermore, by relying on a large FeFET ($4\mu m\times2\mu m$) with a large gate capacitance to represent the LSBs, and large I/O transistors ($L = 0.5\mu m$), the area of the synapse is not reduced despite a lower device count. Finally, while ongoing efforts aim to improve endurance, the endurance of current FeFETs is \texttildelow$10^{5}$ write cycles~\cite{dunkel2017fefet}, which limits the applicability of this design for in-situ training.

\section{Hybrid FeMFET-CMOS Synapse Circuit}
\label{sec:synapse}

\subsection{Synapse circuit design}

\begin{figure}[t]
        \centerline{\includegraphics[width=\columnwidth]{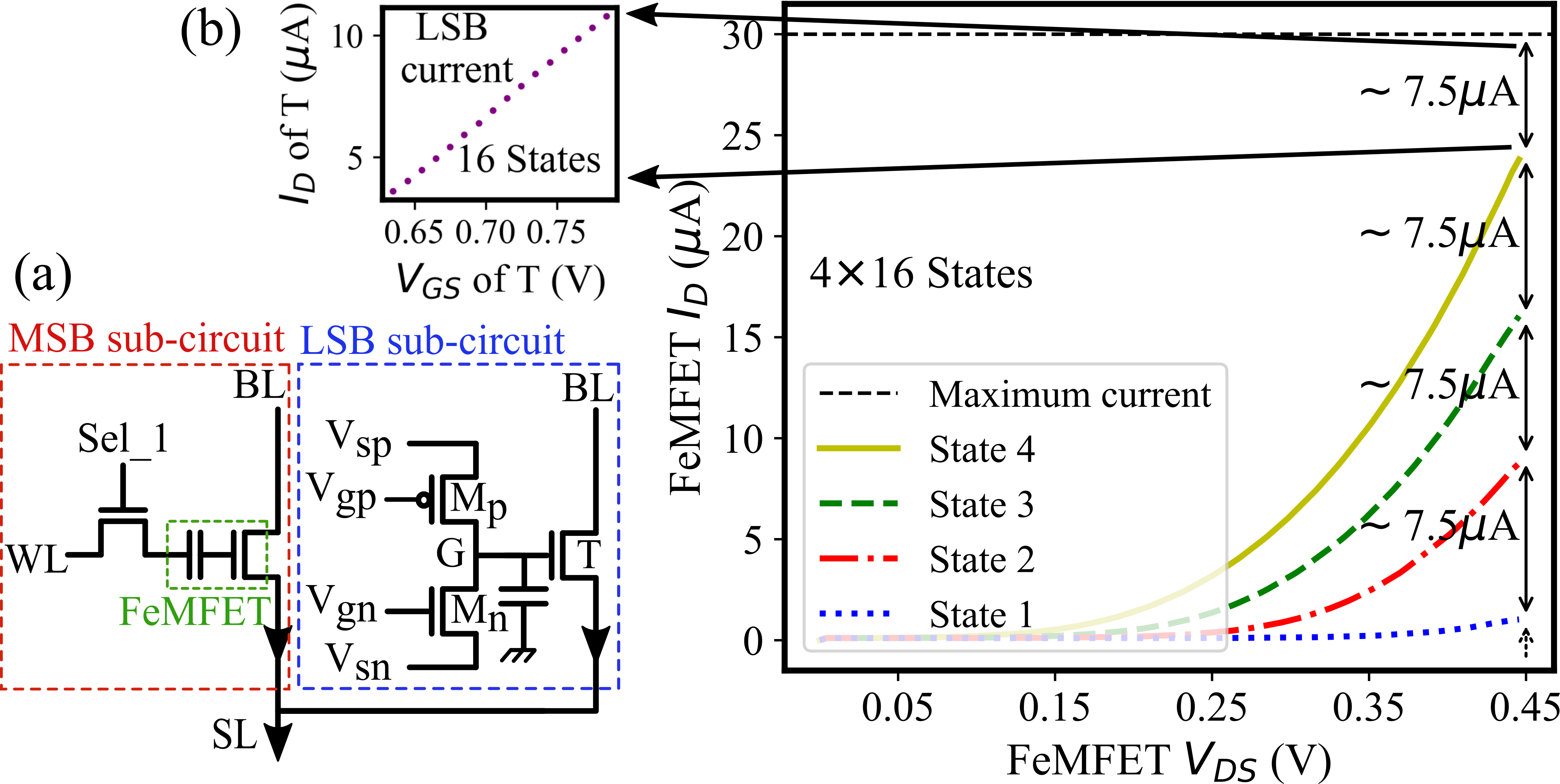}}
        \caption{{\bf (a)} Schematic diagram of the 6-bit synapse; {\bf (b)} total current of the 6-bit synapse, and current from the MSB and LSB sub-circuits.}\label{fig:synapse_6bit}
        \vspace{-0.2in}
\end{figure}

Our proposed 6-bit synapse, (Fig.~\ref{fig:synapse_6bit}(a)) is comprised of 3T1C LSB and 1T1FeMFET MSB sub-circuits. Though our synapse circuit is similar to the 2PCM+3T1C design in Fig.\ref{fig:prior_work}(a), it operates with a single-phase read-out and does not require additional circuitry for arithmetic operations (to be elaborated below). Furthermore, our design reduces the number of elements compared to the 2PCM+3T1C design while attaining the same bit precision. One might wonder if one could substitute a FeFET with a FeMFET in the 2T1F design (Fig.\ref{fig:prior_work}(b)) to alleviate high programming voltages and large I/O transistor overheads. Unfortunately, a simple drop-in replace does not suffice. By changing FeFET polarization, the threshold voltage is changed without altering the shape of the memory window~\cite{jerry2018ferroelectric}, which facilitates the 2T1F design. However, when programming a FeMFET to a different state, both the threshold voltage and the memory window shape change~\cite{ni2018soc}, which prohibits a 2T1FeMFET design.

\begin{figure}
  \begin{subfigure}{\columnwidth}
    \centering
    \includegraphics[width=.95\columnwidth]{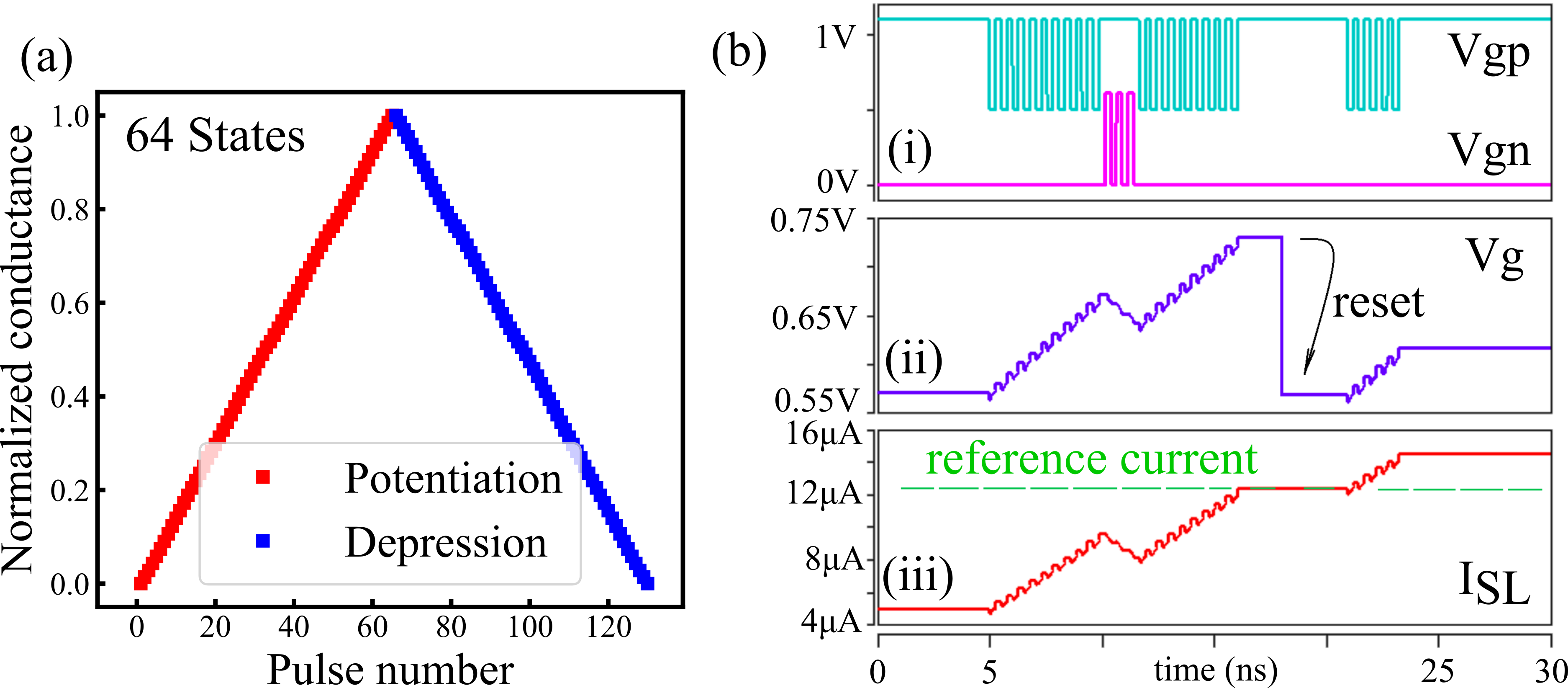}
    \end{subfigure}
    \caption{{\bf (a)} The conductance update curve of the proposed 6-bit synapse, which shows high symmetry and linearity; {\bf (b)} Operation of the proposed 6-bit synapse. Up/down pulses in (b-i) are applied to $V_g$ in (b-ii). Once the current I$_{SL}$ surpasses the reference current in (b-iii), the MSB must be programmed to a higher state and V$_G$ must be reduced to keep I$_{SL}$ the same.}
    \label{fig:condutance_sim_update}
    \vspace{-0.2in}
\end{figure}

Our proposed synapse works as follows. The MSB sub-circuit encodes data as the FeMFET polarization state. Four distinct FeMFET states were chosen as the MSB states which offer current values with increments of $\sim7.5\mu A$, when $V_{DS}$ = 0.45V (see Fig.~\ref{fig:femfet}(b)). The current gap between two consecutive states is filled by the currents of the LSB sub-circuit. Specifically, the LSB sub-circuit encodes data via the current levels obtained from transistor T (Fig.~\ref{fig:synapse_6bit}(a)). Transistors M$_p$ and M$_n$ are sized  such that \textbf{(i)} each positive/negative voltage pulse applied to the gate of M$_p$/M$_n$ changes the voltage of node G (V$_G$) by $\pm$10mV, and \textbf{(ii)} the gate voltage of T lies within a region of 0.57V\textless$V_{GS}$\textless0.73V. This LSB sub-circuit can thus encode 16 states, i.e., each segment of the four colored segments in Fig.~\ref{fig:synapse_6bit}(b) represents 16 states. The combined MSB and LSB sub-circuits allow a single-phase read-out. The difference between the highest and the lowest LSB state currents is $\sim7.2\mu A$, allowing the synapse circuit to generate non-overlapping current values. 
Using the FeMFET model and the 65nm CMOS PTM~\cite{cao2009predictive} model, we have simulated the synapse circuit in SPICE. Fig.~\ref{fig:condutance_sim_update}(a) shows the simulated conductance update curve of the proposed 6-bit synapse, which exhibits high linearity and up/down symmetry.

Fig.~\ref{fig:condutance_sim_update}(b) shows the weight update operation (as wave-forms obtained from SPICE simulation) of the proposed 6-bit synapse circuit. When positive/negative pulse inputs are applied to $V_{gp}$/$V_{gn}$ (Fig.~\ref{fig:condutance_sim_update}(b-i)), V$_G$ increases/decreases by 10mV per pulse. When V$_G$ surpasses 0.73V (Fig.~\ref{fig:condutance_sim_update}(b-ii)), the total current of the synapse (I$_{SL}$) becomes larger than the reference current (Fig.~\ref{fig:condutance_sim_update}(b-iii)), which triggers a weight transfer from the LSB sub-circuit to the MSB sub-circuit. The MSB device must be programmed to a higher state and V$_G$ must be reduced to the voltage of the lowest state, keeping the total current I$_{SL}$ the same. Similarly, if the current drops below the reference current, the MSB must be programmed to a lower state and V$_G$ must be increased to the voltage of the highest state. For the 6-bit design, 3 reference currents are required to distinguish between the 4 FeMFET states.

\subsection{Training and inference with the proposed synapse circuit}
\label{sec:synapse_training}
When performing neural network inference with our synapse, only the MSB sub-circuit is active and its conductance is multiplied with the input voltages to generate outputs. When training neural networks with the proposed synapses, update pulses are applied to the volatile, highly symmetric, and fast LSB sub-circuit to attain high accuracy and rapid training. For every training batch, errors are backpropagated using stochastic gradient descent and appropriate up/down pulses are applied to the LSB. After every \textit{N} (e.g. 100, 200, or 300) batches, the information in the LSB sub-circuit must be transferred to the MSB sub-circuit to \textbf{(i)} preserve information in non-volatile MSBs and \textbf{(ii)} avoid LSB saturation. Determining transfer frequency and what state the LSB should retain after transfer is critical to the training accuracy (see Sec.~\ref{sec:results}).

To elaborate, note that the state of the LSB sub-circuit is degraded as V$_G$ leaks over time. Hence, information cannot be stored in it for long periods. However, information transfer from the LSBs to the MSBs is an expensive operation as \textbf{(i)} the current of the synapse must be examined to ensure that an MSB update is indeed required, and \textbf{(ii)} longer and higher amplitude pulses must be employed to update the MSBs. Thus, information should be transferred to the non-volatile MSB sub-circuit at a rate that avoids information loss of the LSBs, and as infrequently as possible. The impact of the transfer interval length will be evaluated in Sec.~\ref{sec:results}.

To accurately implement weight transfer, the residual information in the LSBs must be preserved. However, this implementation requires {\em additional\/} high-resolution DAC/ADC pairs to program the LSB according to the residual information. To reduce this transfer overhead, once the transfer is conducted, our design simply sets the state of the LSB to its mid-range. Assuming the synapse design with three reference currents $I_1$, $I_2$, and $I_3$, three scenarios can occur after a weight transfer: \textbf{(i)} $I_1 \ll I_{SL} < I_2$, i.e., the synapse is closer to the next MSB state. After the transfer, V$_G$ being programmed to its mid-range state leads to a lower LSB state. \textbf{(ii)} $I_1 < I_{SL} \ll I_2$ (the opposite of case \textbf{(i)}). In this case, the LSB state is higher than it should be. \textbf{(iii)} LSBs are (ideally) in the mid-range state. Clearly, the first two cases incur some loss in the LSB state and reduce the achieved training accuracy. We will evaluate the effects of this in Sec.~\ref{sec:results}.

\begin{figure}[t]
        \centerline{\includegraphics[width=\columnwidth]{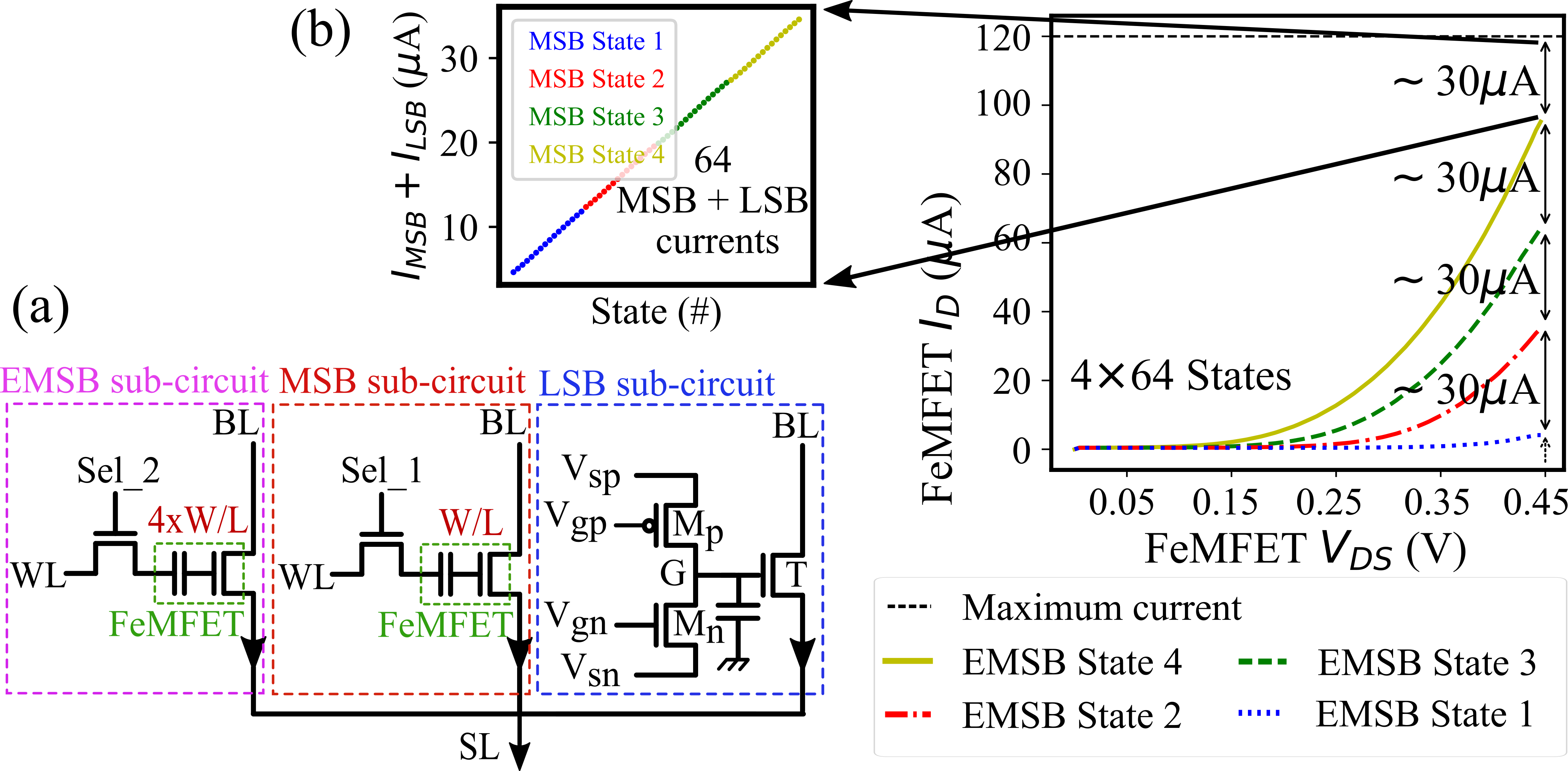}}
        \caption{\textbf{(a)} Schematic diagram of the 8-bit synapse; \textbf{(b)} total current of the 8-bit synapse, and current from the EMSB, MSB, and LSB sub-circuits.}\label{fig:synapse_8bit}
        \vspace{-0.2in}
\end{figure}

\subsection{8-bit extension of the proposed synapse circuit }
To improve the accuracy in both training and inference for more complicated datasets such as CIFAR10, we propose to extend the design in Fig.~\ref{fig:synapse_6bit}(a) to an 8-bit synapse circuit. Specifically, an extended MSB (EMSB) sub-circuit is added to the 6-bit circuit as shown in Fig.\ref{fig:synapse_8bit}(a). The total current of the 8-bit circuit with two MSB (EMSB  and MSB) sub-circuits is shown in Fig.~\ref{fig:synapse_8bit}(b). The W/L of the FeMFET in the EMSB sub-circuit is increased by 4\texttimes~when compared to that of the MSB sub-circuit to allow more distinct conductance states. Similar to the 6-bit design, in this circuit, the difference between the states of the EMSB sub-circuit is filled by the MSB current values, and those of the MSBs are filled by the LSBs, to realize 8-bit precision. The weight update operation of this design is similar to the 6-bit design, with the difference of having 3 extra reference currents to distinguish EMSB device states.

\section{Evaluation}
\label{sec:results}

We first evaluate the training accuracy of our synapses. We train a variant of LeNet with the MNIST~\cite{lecun-mnisthandwrittendigit-2010} dataset using the 6-bit synapse circuit, and a VGG-like network with the CIFAR-10 dataset using the 8-bit design. The LeNet and VGG networks are identical to the networks trained in~\cite{sun2018exploiting}, hence the results can be directly compared. The LeNet(VGG) network has 2(6) convolutional and 2(2) fully connected layers. We model the characteristics of our synapse circuits with TensorFlow~\cite{abadi2016tensorflow}. We use gradient descent, and choose a batch size of 100. We evaluate the effect of different weight transfer intervals on achievable accuracy. We further evaluate our 6-bit synapse circuit using the NeuroSim+~\cite{chen2017neurosim+} tool and compare it with other hybrid synapse circuits. We also present benchmarking results on area and leakage power.

\subsection{Neural network training accuracy evaluation}
\label{sec:training}

\subfile{Figs/mnist_transfer}

Fig.~\ref{fig:mnist_transfer} shows the results of training the LeNet network with the 6-bit synapse circuit with MNIST. The software baseline is a network trained using ideal linear 6-bit weights. The ``synapse - Ideal" data point shows the accuracy of a network trained with our synapse circuit, assuming that weight transfer occurs when the LSB is saturated and no residual information is lost on the LSB after transfer. The achieved accuracy is \texttildelow98.5\% (\texttildelow0.1\% lower than the baseline). Recall that weight transfers may cause LSB information loss (Sec.~\ref{sec:synapse_training}). Thus, shorter transfer intervals lead to the accumulation of information loss on the LSB due to more frequent transfers. Hence, the achieved accuracy of our synapse is directly correlated with the transfer interval. A transfer interval of 300 batches achieves \texttildelow98.2\% accuracy, with only \texttildelow0.4\% degradation compared to the baseline, whereas a transfer interval of 100 shows \texttildelow1.6\% degradation and achieves a \texttildelow97\% accuracy.

Though, theoretically, longer transfer intervals lead to higher accuracy, the existence of the leakage current path in the LSB sub-circuit results in LSB information decay. Hence, to choose a suitable transfer interval, we estimate the required time for training LeNet (forward pass + backpropagation + weight update) on a single batch (batch size is 100) using the proposed synapse circuits. With 250ps pulses and the same array assumptions as~\cite{sun2018exploiting}, i.e., 128$\times$128 array size, 2ns read delay, and 8 columns sharing an ADC, we find this time to be \texttildelow700ns. We then evaluate the leakage of node G in Fig.~\ref{fig:synapse_6bit}(b) and find the time for V$_G$ to drop 10mV (equivalent to one LSB state) to be 215$\mu$s in the worst-case scenario. This allows for a batch transfer interval of \texttildelow300 when training LeNet with the MNIST dataset, whereas the 2T1F design can only achieve a transfer interval of \texttildelow200~\cite{sun2018exploiting}. Comparing our accuracy results with the 2T1F design shows an improvement of almost \texttildelow1\% in accuracy due to both increased transfer interval and a more linear and symmetric update curve (Fig.~\ref{fig:condutance_sim_update}(a)). 

When considering a VGG network, an 8-bit synapse, the CIFAR-10 dataset, and a transfer interval of 300 batches, the achievable accuracy is 89.3\% -- just 0.3\% lower than a baseline software implementation of 89.6\%. Having 4 bits for MSBs reduces the significance of the 4 LSBs compared to 2 bits for MSBs. Also, LSB state loss during transfers adds stochasticity to the weights. However, as a larger network must be trained, the training time per batch increases by \texttildelow3\texttimes~when compared to LeNet. Hence, we can only use a transfer interval of 100 batches, which achieves an accuracy of \texttildelow88.8\% -- 1.8\% higher than the 2T1F design (87\%). The size of the network is not considered when training VGG in~\cite{sun2018exploiting} and the transfer interval for LeNet is assumed. Again, higher bit precision and a more linear/symmetric weight update curve improve accuracy.

\subsection{System-level benchmark results}

\begin{table}[t]
\renewcommand{\arraystretch}{1.3}
\caption{Device characteristics and system-level benchmark results of the 6-bit hybrid synapse designs (65nm node).}
\label{table:benchmarks}
\resizebox{\columnwidth}{!}{%
\begin{tabular}{|l|c|c|c|}
\hline
Synapse                                                                & 2PCM+3T1C~\cite{ambrogio2018equivalent}                                                                        & 2T1F~\cite{sun2018exploiting}                                                                       & proposed synapse                                                          \\ \hline \hline
LSB update pulse                                                          & 1V/300ps                                                                                & 1V/300ps                                                                          & 1.1V/250ps                                                                         \\ \hline
MSB update pulse                                                          & 0.7V (avg)/6$\mu s$                                                                                & 2-4V/3$\mu s$                                                                          & 1.4-1.8V/100ns                                                                         \\ \hline
MSB endurance                                                          & 10$^8$                                                                           & 10$^5$                                                                     & 10$^{10}$                                                                   \\ \hline
Area ($mm^2$)                                                       & 2.65                                                                          & 2.73                                                                    & 1.96                                                                   \\ \hline
Leakage power (mW)                                                     & 7.98                                                                             & 14.46                                                                      & 7.98                                                                      \\ \hline
MNIST accuracy (\%)                                                     & 97.95                                                                             & 97.3                                                                      & 98.2                                                                      \\ \hline
\end{tabular}
}
\vspace{-0.2in}
\end{table}

System-level benchmark results of the 2T1F and 2PCM+3T1C hybrid synapse circuits are presented in~\cite{luo2019benchmark} using the NeuroSim+~\cite{chen2017neurosim+} tool. For fair comparison, we benchmark our proposed 6-bit synapse circuit with the same assumptions made in~\cite{luo2019benchmark}. Table~\ref{table:benchmarks} presents the device parameters as well as the area and leakage power for training an MLP with 400\texttimes200\texttimes10 neurons with MNIST. The proposed 6-bit synapse circuit reduces the area by 26\%, given the reduced number of devices when compared to the 2PCM+3T1C design. As the size of the employed transistors in the 2T1F design is large,  smaller transistor sizing afforded by our FeMFET approach reduces the leakage power of our design by 44.8\%. Update pulse speed is improved by 16.7\% via tuning of the LSB sub-circuit. Furthermore, FeMFETs yield 2 and 5 orders of magnitude more endurance cycles compared to PCMs~\cite{ambrogio2018equivalent} and FeFETs\cite{dunkel2017fefet}, respectively, which is favorable when using the circuits for in-situ training.

\section{Conclusion}
In this paper, new hybrid FeMFET-CMOS analog synapse circuits offering 6-bit and 8-bit precision for in-situ training of neural networks were proposed. Our design is superior to other hybrid synapse designs in terms of area, power, performance, and endurance, and approaches software accuracies.

\section*{Acknowledgment}
\footnotesize{This work was supported by ASCENT, one of the six SRC/DARPA JUMP centers under task ID 2776.043.}

\newpage
\IEEEtriggeratref{9}
\bibliographystyle{IEEEtran}
\bibliography{bibfile}

\end{document}